# Metal-organic Frameworks: Possible New Two-Dimensional Magnetic and Topological materials

Jie Li and Ruqian Wu*

*Department of Physics and Astronomy, University of California, Irvine, California 92697-4575, USA.*

Finding new two-dimensional (2D) materials with novel quantum properties is highly desirable for technological innovations. In this work, we studied a series of metal-organic frameworks (MOFs) with different metal cores and discovered various attractive properties, such as room-temperature magnetic ordering, strong perpendicular magnetic anisotropy, huge topological band gap (>200meV), and excellent spin-filtering performance. As many MOFs have been successfully synthesized in experiments, our results suggest realistic new 2D functional materials for the design of spintronic nanodevices.

* E-mail: wur@uci.edu.

# 1. Introduction

Ever since the successful synthesis of graphene,[1] two-dimensional (2D) materials have attracted tremendous research interest. A large verity of 2D materials have been predicted, fabricated and characterized, such as hexagonal boron nitride (h-BN),[2] silicone,[3] germanene,[4] transition metal dichalcogenides,[5] black phosphorous,[6] $Cr_2Ge_2Te_6$,[7] and $CrI_3$.[8] They all have exotic quantum properties such as the quantum spin Hall effect (QSHE), quantum anomalous Hall effect (QAHE), valley-polarized anomalous Hall effect (VAHE), high carrier mobility, or stable low-dimensional ferromagnetic ordering. Searching for functional 2D materials with emergent physical properties is currently at the forefront of research activities in several subareas, including condensed mater physics, chemistry, nanoscience, and materials science.

Among potential 2D materials in this realm, metal-organic frameworks (MOFs) are particularly attractive as they have various advantages such as easy fabrication and manipulation, high mechanical flexibility, and low cost. Many stable MOFs haves been predicted through theoretical studies, such as triphenyl-lead based topological 2D materials,[9-10] nickel bis complex π-nanosheet,[11] π-conjugated covalent-organic frameworks,[12] and many of them have already been synthesized in experiments. For example, Shi et al. synthesized a series of MOFs on the Au(111) surface by positing tripyridyl ligands with transition metal atoms.[13] Pawin et al. synthesized an anthracenedicarbonitrile based coordination network on the Cu(111) surface.[14] Koudia et al. synthesized 2D transition-metal phthalocyanine-based MOFs by annealing Pc molecules on the Ag(111) surface.[15] Abdurakhmanova et al. synthesized and explored TM-TCNQ networks.[16-18] Obviously, these successes have paved the way for the design of new 2D functional MOFs with distinct properties that are desired for applications.

In this letter, we theoretically predict a series of new 2D MOFs with a simple hexagonal lattice based on transition metal tris (dithiolene) complexes which have been recently synthesized in experiments.[19-20] We explore their functionalities such as magnetic ordering and topological features through systematic ab-initio calculations. As we adjust the species of the metal cores, these 2D MOFs manifest various attractive properties such as high Curie temperature, strong out-of-plane magnetic anisotropy, sizeable topological band gap (>200meV) and excellent spin-filtering performance. These findings suggest the suitability of these 2D MOF materials for the design of innovative nano-devices.

## 2. Calculation methods

All density functional theory (DFT) calculations were carried out using the Vienna ab-initio simulation package (VASP) with the projector augmented wave (PAW) method was adopted for the interaction between valence electrons and ionic cores,[21-22] and the energy cutoff for the plane wave basis expansion was set to 700 eV. The spin-polarized generalized-gradient approximation (GGA) with the functional developed by Perdew-Burke-Ernzerhof (PBE) was choosed for the exchange-correlation functional.[23] The vdW correction (DFT-D3) was included for the description of dispersion forces.[24] To sample the two-dimensional Brillouin zone, we used a 9×9 k-grid mesh. All atoms were fully relaxed using the conjugated gradient method for the energy minimization until the force on each atom became smaller than 0.01 eV/Å. The edge states of 2D TM-Hex were calculated by TB model based on the maximally localized Wannier functions (MLWFs) as implemented in Wannier90 code [25] and Wannier tools code.[26]

# 3. Results and discussion

As shown in Fig. 1(a), these 2D MOF materials take a honeycomb lattice and each sublattice site has a molecule with three ligands around its vanadyl core. The ligand size can be changed, for example, among structures I, II and III as depicted in Fig. S1.[19-20] In this work, we designed two 2D metal-organic hexagonal lattice (M-Hex) with different porosity as shown in Fig. 1(a) and Fig. S2 (label as M-Hex-I and M-Hex-II, respectively). We tested different transition metal cores, including all 3d, 4d and 5d transition metal atoms. The 2D planar structure was found to maintain well after the structural relaxation procedures. The optimized lattice constants of a few selected systems are shown in Fig. S3.

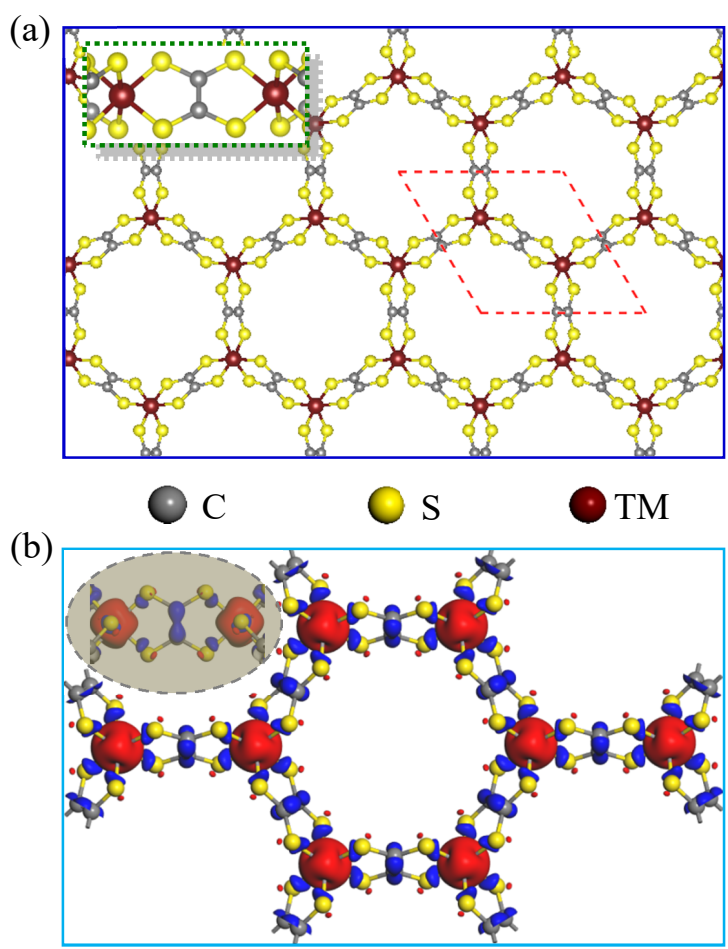

Fig. 1. (a) The top and side (inset) views of schematic structure of the predicted 2D M-Hex-I (the dashed square shows the supercell). (b) The top and side (inset) views of spin density of Mn-Hex-I

*(red and blue color represent positive and negative spin densities, respectively.).*

To further confirm their thermal and dynamic stability, we took Mn (Mn-Hex-I/II), Re (Re-Hex-I) and Os (Os-Hex-I) as examples and performed phonon and molecular dynamics simulations. The corresponding phonon bands are shown in Fig. S4 in the Supplementary Information. The absence of imaginary frequency branch indicates that these systems are dynamically stable. Furthermore, we kept them at 300K for 10ps (5000 steps) through ab initio molecular dynamics (AIMD) simulations with a 5×5 supercell. The AIMD results show that the total energies fluctuate around their equilibrium values without noticeable sudden change (see Fig. S5 in the Supplementary Information). As no structure destruction was observed, we believe that these lattices are thermally stable at least up to room temperature. In the following, we separate our discussions in two parts: systems with either sustainable magnetization (3d core atoms) or with strong spin orbit coupling (4d and 5d core atoms). The former focuses on the magnetic properties and the latter focuses on the topological properties.

From our DFT calculations, only (V, Cr, Mn, Fe)-Hex-I(II) MOFs are magnetic, as shown in Fig. S3 in the Supplementary Information. For instance, each Mn atom in the Mn-Hex-I/II lattice has a charge state of +3 and a spin magnetic moment of 2.0 $u_B$. As seen from the spin density in Fig. 1(b), the adjacent ligands also have an antiparallel spin polarization but the magnet moments are negligible, < 0.01 $u_B$/atom. Importantly, Mn(V, Fe)-Hex-I(II) lattices prefer ferromagnetic (FM) coupling. The corresponding exchange parameters, $J_i$, are obtained by mapping the DFT total energies of different magnetic configurations (see Fig. S6 in the Supplementary Information) to the classical Heisenberg Hamiltonian:

$$H = H_0 - J_1 \sum_{<i,j>} S_i \cdot S_j - J_2 \sum_{<i,j>} S_i \cdot S_j \qquad (1)$$

where $J_1$, and $J_2$ represent the nearest and the next nearest neighbor exchange interactions, respectively, (see Table SI in the Supplementary Information). Here, we only consider $J_1$ and $J_2$, since the third nearest neighbor exchange interactions are negligible due to the large distance. As shown in Table SI, the exchange parameters are unexpectedly large for the porous networks, indicating possible high Curie temperatures for them. In addition, Mn-Hex-I (II) also have out-of-plane spin orientation and large magnetic anisotropic energy (MAE) of 0.75 (0.86) meV. Using the torque method, the corresponding Fermi level dependent total and spin channel decomposed MAEs are shown in Fig. S7 in the Supplementary Information. One may see that their large MAEs in a broad energy range around the actual Fermi level are mostly from the cross-spin contributions. With the DFT results of exchange and magnetic anisotropy energies, we calculate their $T_c$ by using the renormalized spin-wave theory (RSWT).[27-28] The renormalized magnetization ($M(T)/M(0)$) as a function of temperature T is shown in Fig. 2(a), and the Curie temperature is determined by the location where the renormalized magnetization drops zero. One may see that these systems have exceedingly high Curie temperatures compared to other 2D vdW magnetic layers (45K, 66K and 160 K for $CrI_3$, $Cr_2Ge_2Te_6$ and CrOCl, respectively).[7,8,29] Especially, $T_c$ of Mn-Hex-I is as high as ~323K. Even though the exchange interactions and Curie temperature decrease as the ligands become longer, the Curie temperature of Mn-Hex-II is still as high as 160 K. The possibility of having magnetic ordering up to such high temperatures in 2D structures makes them very attractive for spintronic applications.

To explore the mechanism of strong exchange coupling and electronic properties in MOFs, the projected density of states (PDOS) of Mn-Hex-I are given in Fig. 2(b), with projections to different atoms (upper panel) and different Slater orbitals (lower panel). First, the MOF is half-metallic, with a large band gap in the majority spin channel. Strong hybrization occurs among Mn, C and S orbitals, as manifested by the large broadenings of the DOS peaks. The involvement of delocalized π electrons of C and S atoms certainly mediate the exchange coupling between transition metal atoms along the ligands, as was discussed by several authors for similar systems.[30-33] Even for M-Hex-II, there are still hybridized states around the Fermi level so the π electrons still effectively assist the exchange coupling in the MOFs, as shown by the charge distribution of states in the range of ±0.2 eV around the Fermi level in Fig. S8 in the Supplementary Information. Note that these states around the Fermi level is 100% spin polarized in the FM state, and hence may mediate the long-range exchange interactions in MOFs, much like an "extended" direct exchange mechanism. In contrast, the hybridization is much weakened in the AFM phase, as the $d_{xy/x^2-y^2}$ and $d_{xz/yz}$ peaks skink their widths (see the dashed lines in the lower panel of Fig. 2(b)). This is an indication that the Mn-ligand hybridization is much reduced and the hopping between adjacent Mn atoms becomes forbidden, which is unfavorable in energy.

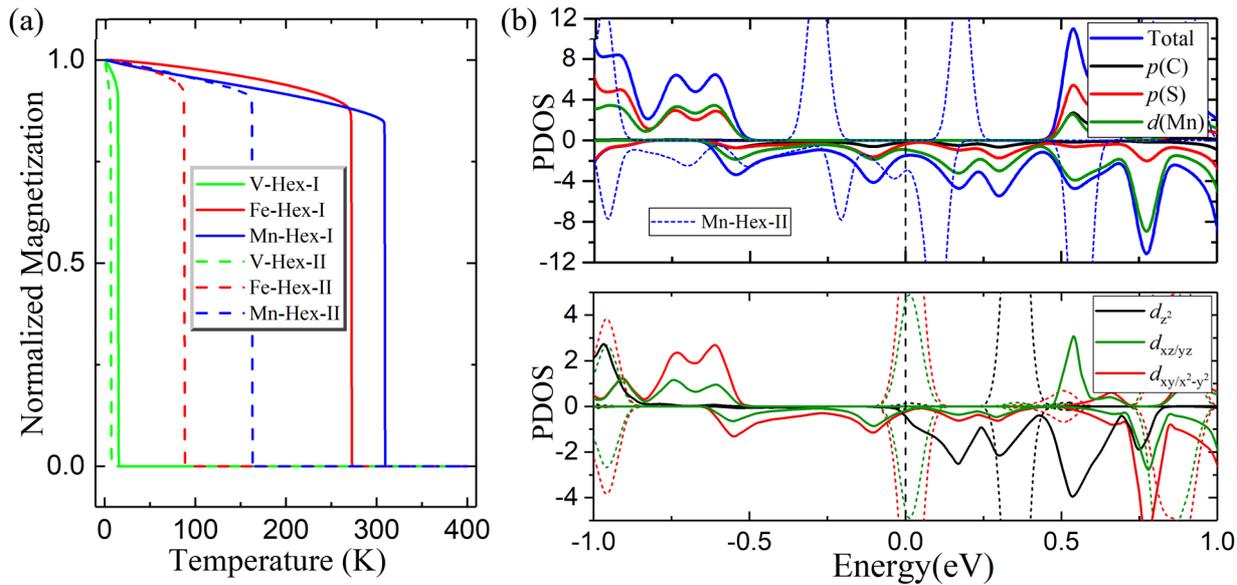

Fig. 2. (a) The renormalized magnetization as a function of temperature T for V(Fe, Mn)-Hex-I(II), respectively. (b) The corresponding PDOS of Mn-Hex-I(II). (Top) the total DOS and the PDOS of

*C, S and Mn atoms of Mn-Hex-I in the FM state, the dash lines are the total DOS of Mn-Hex-II. (Down) the PDOS of d orbitals of Mn atoms of Mn-Hex-I in the FM state (solid lines) and AFM state (dash lines).*

As we also look for possible nontrivial topological properties from these honeycomb MOFs, the band structures of Mn-Hex-I without and with spin orbit coupling (SOC) are given in Fig. 3. Indeed, Mn-Hex-I has the Dirac cones at the K and K` points, but there is another band that cross the Fermi level in the vicinity around the Γ point. Therefore, Mn-Hex-I is a half-metal with a large band gap in the majority spin channel, 1.07 eV (0.42 eV for Mn-Hex-II). As the FM ordering may sustain at high temperature, one possible use of these exotic porous 2D magnetic monolayers is making spin filters. To this end, we investigated two prototypical Mn-Hex-I nanoribbons, i.e., with zigzag edges (4 zigzag chains in width) or armchair edges (3 armchair chains in width). As shown in Fig. S9 in the Supplementary Information, only the spin down channel of these ribbons opens for electron conduction in a broad energy range and hence they can be excellent for spin-filtering applications. Furthermore, the armchair Mn-Hex-I nanoribbon has a small gap (<0.1 eV) in the majority spin channel and is expected to be highly responsive to gating voltages and can be used for switching as well. Note that Mn-Hex-I has two other topological nontrivial gaps (13.9 meV and 24.4 meV) as shown in the dash boxes in Fig. 3(b), which are relatively large compared to several previous predictions of MOFs,[34-36] but they are too far away from the Fermi level. Interestingly, Mn-Hex-II has the Dirac cones right at the Fermi level, with a SOC induce gap of 3 meV (cf. Fig. S10). Although this gap is small, it might be worthwhile to try for the realization of the quantum anomalous Hall effect.

All MOFs with 5d cores are nonmagnetic and thus it is interesting to see if they can manifest

QSHE, as they have strong SOC at the cores. From the band structures without and with SOC in Fig. 4(a), we found that Os-Hex-I has a topologically nontrivial band gap right at the Fermi level. Without the SOC term in calculations, the highest valence band and the lowest conduction band touch each other around the Γ point. With SOC, a huge gap of about 225.6 meV opens and the Fermi level naturally lies in the gap, a desirable feature for topological materials. We also calculated the corresponding band structures by switching off SOCs of carbon, sulfur and osmium atoms. As shown in Fig. S11, the gap changes to 7 meV without SOC of Os and to 231 meV without SOC of C and S, respectively. Obviously, the large gap mostly results from the SOC of Os. There are also Dirac cones for Os-Hex-I, like graphene, but they are at -0.5 eV. These Dirac cones are accessible when Os is replaced by Re, which reduces one electron from the unit cell, as shown in Fig. S12 in the Supporting Information.

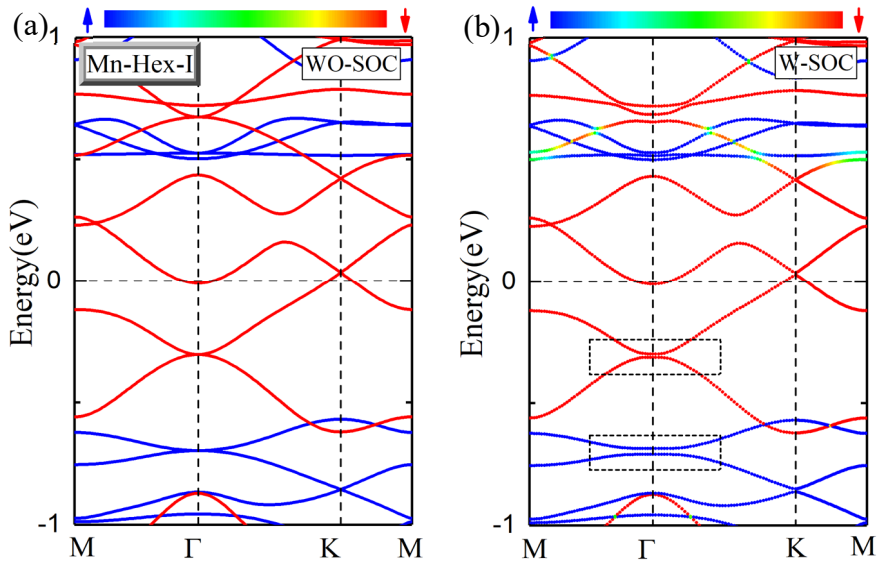

Fig. 3. (a) (b) The band structures of Mn-Hex-I without and with SOC, respectively.

To verify if the SOC induced band gaps of Os-Hex-I and Re-Hex-I are topologically nontrivial, their $Z_2$ numbers and edge states were calculated. As shown in Fig. 4(b), we may see that $Z_2=1$ for Os-Hex-I by counting the positive and negative n-field numbers over half of the torus. This undoubtedly indicates that Os-Hex-I is a strong 2D topological insulator. Similarly, Re-Hex-I is a strong TI material as well, as shown in Fig. S12 in the Supplementary Information. As the band structure of Re-Hex-I is similar with that of graphene but with a much larger topological band gap (80 meV), it offers a new platform for studies of QSHE in honeycomb lattice. Obviously, the high structural stability and large topological gaps of these MOFs are very promising for practical applications and deserve experimental verifications.

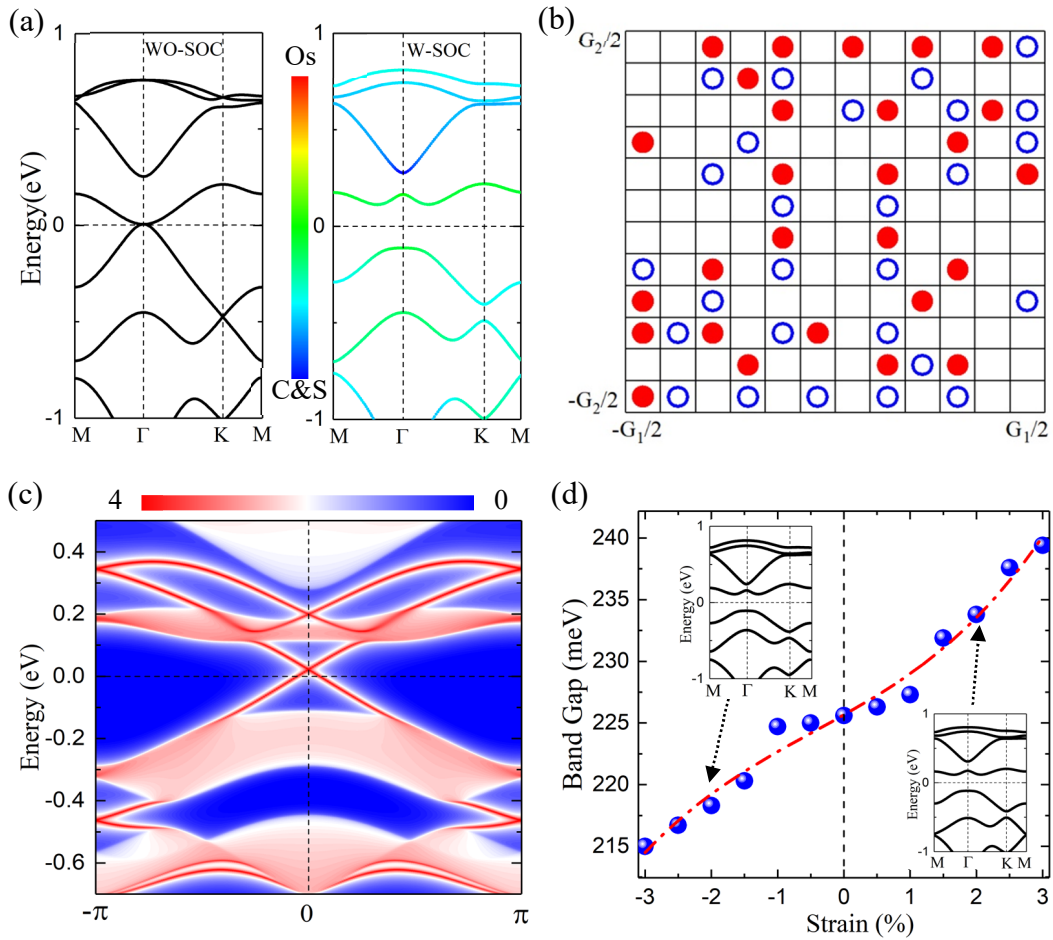

Fig. 4. (a) The band structure of Os-Hex-I without and with SOC, respectively. Color bar indicates the atom-projected weights. (b) The n-field configuration with red solid, blue hollow circles and blank boxes denoting n=-1, n=1 and n=0. (c) The corresponding 1D band structure and edge states. (d) SOC-induced band gap as a function of in-plane strain.

As further evidence, we constructed an Os-Hex-I nanoribbon (~100nm in width) and see if it has the topologically protected edge states. To handle large number of atoms in the system, we used a tight-binding (TB) model with parameters obtained by fitting the DFT band structure of 2D Os-Hex-I. In the TB model, we used s and p (s and d) orbitals as the basis for C and S (Os) atoms. The TB band structure (see Fig. S13 in the supporting Information) agree with DFT band structure very well, showing the high quality of our fitting. In Fig. 4(c), one may see two bands from edge states across the Fermi level. They connect the bulk conduction and valence bands and present in the bulk-like band gap. Again, this supports the conclusion that Os-Hex-I MOF is a strong 2D topological insulator and can be used for the realization of the quantum spin Hall effect.

Flex is a common way for the manipulation of electronic properties of 2D materials, especially for engineering the topological phases.[37-39] Here, we considered the effect of a biaxial strain on the band structure of Os-Hex-I (see in Fig. 4(d)). One may see that Os-Hex-I can hold a large gap ( > 210 meV) with a similar band structures in a reasonable range of strain (-3% ~3%).

## 4. Conclusions

In summary, we proposed several new 2D functional materials based on honeycomb metal-organic frameworks, and investigated their structural, magnetic, electronic, and topological properties. DFT calculations showed that MOFs with Mn has a strong magnetization up to room temperature and can be used for switching and spin-filtering applications. On the other hand, MOFs with Os and Re cores are strong 2D topological materials with robust topologically protected edge states in their ribbons. As they may have a gap larger than 200 meV, they are very promising for the realization of quantum spin Hall effect at high temperature. Our studies suggest a new and reliable strategy for the design of functional spintronic and topotronic materials.

## Conflicts of interest

The authors declare no competing financial interest.

## Acknowledgements

Work was supported by US DOE, Basic Energy Science (Grant No. DE-FG02-05ER46237). Calculations were performed on parallel computers at NERSC.

## References

1 K.S. Novoselov, A.K. Geim, S.V. Morozov, D. Jiang, Y. Zhang, S.V. Dubonos, I.V. Grigorieva, and A.A. Firsov, *Science*, 2004, **306**, 666.

2 C.R. Dean, A.F. Young, I. Meric, C. Lee, L. Wang, S. Sorgenfrei, K. Watanabe, T. Taniguchi, P. Kim, K.L. Shepard, and J. Hone, *Nat. Nanotechnol.*, 2010, **5**, 722.

3 S. Cahangirov, M. Topsakal, E. Aktürk, H. Şahin, and S. Ciraci, *Phys. Rev. Lett.*, 2009, **102**, 236804.

4 L. Li, S. Z. Lu, J. Pan. Z. Qin, Y.Q. Wang, Y. Wang, G.Y. Cao, S. Du, and H.J. Gao, *Adv. Mater.*, 2014, **26**, 4820.

5 Q.H. Wang, K. Kalantar-Zadeh, A. Kis, J.N. Coleman, and M.S. Strano, *Nat. Nanotechnol.*, 2012, **7**, 699.

6 H. Liu, A.T. Neal, Z. Zhu, Z. Luo, X. Xu, D. Tománek, and P.D. Ye, ACS Nano, 2014, **8**, 4033.

7 C. Gong, L. Li, Z. Li, H. Ji, A. Stern, Y. Xia, T. Cao, W. Bao, C. Wang, Y. Wang, Z.Q. Qiu, R.J. Cava, S.G. Louie, J. Xia, and X. Zhang, *Nature*, 2017, **546**, 265.

8 B. Huang, G. Clark, E. Navarro-Moratalla, D.R. Klein, R. Cheng, K.L. Seyler, D. Zhong, E. Schmidgall, M.A. McGuire, D.H. Cobden, W. Yao, D. Xiao, P. Jarillo-Herrero, and X. Xu, *Nature*, 2017, **546**, 270.

9 Z.F. Wang, Z. Liu, and F. Liu, *Nat. Commun.*, 2013, **4**, 1471.

10 H.J. Kim, C. Li, J. Feng, J.H. Cho, and Z. Zhang, *Phys. Rev. B*, 2016, **93**, 041404 (R).

11 T. Kambe, R. Sakamoto, T. Kusamoto, T. Pal, N. Fukui, K. Hoshiko, T. Shimojima, Z. Wang, T. Hirahara, K. Ishizaka, S. Hasegawa, F. Liu, and H. Nishihara, *J. Am. Chem. Soc.*, 2014, **136**, 14357.

12 L. Dong, Y. Kim, D. Er, A.M. Rappe, and V.B. Shenoy, Phys. Rev. Lett., 2016, **116**, 096601.

13 Z. Shi, J. Liu, T. Lin, F. Xia, P.N. Liu, and N. Lin, *J. Am. Chem. Soc.*, 2011, **133**, 6150.

14 G. Pawin, K.L. Wong, D. Kim, D. Sun, L. Bartels, S. Hong, T.S. Rahman, R. Carp, and M. Marsella, *Angew. Chem. Int. Ed.*, 2008, **120**, 8570.

15 M. Koudia, and M. Abel, *Chem. Commun.*, 2014, **50**, 8565.

16 N. Abdurakhmanova, T.C. Tseng, A. Langner, C.S. Kley, V. Sessi, S. Stepanow, and K. Kern, *Phys. Rev. Lett.*, 2013, **110**, 027202.